\input xy
\xyoption{all} 
\input epsf.tex

\documentclass[twocolumn,showpacs,preprintnumbers,amsmath,amssymb]{revtex4}

\usepackage{graphicx}
\usepackage{dcolumn}
\usepackage{bm}
\usepackage{latexsym}
\usepackage{mathrsfs}

\begin{document}

\title{Closed String Tachyon: Inflation and Cosmological Collapse}

\author{Celia Escamilla-Rivera}
\email{Celia.EscamillaRivera@astro.ox.ac.uk}
\email{celia_escamilla@ehu.es}
\affiliation{Department of Astrophysics, University of Oxford, DWB,
Kebble Road, Oxford, OX1 3RH, UK}
\affiliation{Fisika Teorikoaren eta Zientziaren Historia Saila, Zientzia 
eta Teknologia Fakultatea, Euskal Herriko Unibertsitatea, 644 Posta 
Kutxatila, 48080, Bilbao, Spain.}
 
\author{Gerardo Garc\'ia-Jim\'enez}
\email{ggarcia@fcfm.buap.mx}
\affiliation{Facultad de Ciencias F\'isico Matem\'aticas, 
Universidad Aut\'onoma de Puebla, P.O. Box 1364, 72000, Puebla, Mexico.} 
 
\author{Oscar Loaiza-Brito}
\email{oloaiza@fisica.ugto.mx}
\affiliation{Departamento de F\'isica, Divisi\'on de Ciencias e Ingenier\'ia, Campus Le\'on, Universidad de
Guanajuato, P.O. Box E-143, 37150, Le\'on, Guanajuato., Mexico.}  
    
\author{Octavio Obreg\'on}
\email{obregon@fisica.ugto.mx}
\affiliation{Departamento de F\'isica, Divisi\'on de Ciencias e Ingenier\'ia, Campus Le\'on, Universidad de
Guanajuato, P.O. Box E-143, 37150, Le\'on, Guanajuato., Mexico.}
 
\date{\today}


\begin{abstract}
By compactifying a critical bosonic string theory on an internal non-flat space  with a constant volume, we study the role played by the closed string tachyon in the cosmology of the effective four-dimensional space-time.  The effective tachyon potential consists on a negative constant related to the internal curvature space and a polynomial with only quadratic and quartic terms of the tachyon field.  Based on it, 
we present a solution  for the tachyon field and the scale factor, which describes an accelerated  universe which expands to a maximum value before collapsing. At early times, the closed string tachyon potential behaves as a cosmological constant driving the Universe to an expansion.  The value of the cosmological constant is determined by the curvature of the internal space which also fixes the value of the vacuum energy.  As time evolves, inflation is present in our models, and it finishes long before the collapsing. At late times, we show that the collapse of the Universe starts as soon as the tachyon field condensates. We also comment on the consistency of our solution at early times at which quantum aspects become important under the perspective of quantum cosmology. Finally we briefly mention the relation among the curvature of the internal space and the value of the energy vacuum in a non-constant internal volume scenario.


\end{abstract}


         
\maketitle


\section{Introduction}

It is known from the early days of string theory that the spectrum of strings 
does contain a negative mass squared vibration called the tachyon.  In superstring theory, the most well 
known scenario concerns the presence of a tachyonic mode on the open string 
spectrum between pairs of D-branes and anti-D-branes.
Since the seminal papers by A. Sen \cite{Sen:1998ii, Sen:1998sm} (see also
\cite{Sen:2002nu,Sen:2002an, Zwiebach:1997fe} for nice reviews and references therein) we now 
understand that a tachyonic mode signals the fact that we are perturbating 
strings in a false vacuum, corresponding to a maximum of the tachyonic 
potential.\\

In order to be in the minimal state of energy, the tachyon rolls down to the minimum of the potential, and the 
perturbative approach of the theory becomes reliable. The tachyon mode disappears 
by acquiring a vacuum expectation value (vev) which exactly cancels the 
original negative squared mass. In the context of a brane-antibrane system, 
the energy absorbed by the tachyon by rolling down the potential is taken 
from the mass-tension of the branes. This process is called tachyon 
condensation.\\

In fact, open string tachyon condensation drives the decay of branes, but also is the 
mechanism by which D-branes can be constructed from a pair (or pairs) of 
higher dimensional brane-antibranes.  So in this sense D-branes can be 
thought of as being made of tachyons. Broadly speaking, condensation 
of open string tachyon yields the disappearance of the D-branes $-$the 
source of the tachyon modes$-$.\\

The analysis of this process has been made in different contexts within 
string theory. One of them concerns  cosmological applications, which 
have generated a huge interest in the community over the last years 
(see for instance \cite{Gibbons:2002md, Gibbons:2003gb, Farajollahi:2011ym, Farajollahi:2011jr}, and references 
therein). The prototypical scenario considers the expansion of the 
Universe on a $D$-brane as an effective model of an open string tachyonic 
mode connecting a pair of brane-antibrane \cite{Burgess:2001fx}. In the same line, the study 
of tachyon condensation in open string theory has led to a number of 
important insights as
the identification of the tachyon field as a proper time in quantum cosmology
\cite{Sen:2002qa, Sen:2002vv} and the fact that this relationship is broken 
in the presence of the electromagnetic field \cite{GarciaCompean:2005zn, EscamillaRivera:2010rw}. However, the problem of time in quantum gravity is still an open question (see for instance \cite{Bojowald:2010qw} and references therein).\\

On the other hand, tachyonic modes are not exclusively related to open strings. Closed 
string tachyon modes arise for instance, in the bosonic string spectrum and at orbifold singularities within the context of superstring theory
 \cite{Adams:2001sv, Vafa:2001ra, Headrick:2004hz, Bergman:2006pd}. 
Open and closed string are quite distinct, and describe disjoint aspects when 
applied to cosmology. At tree level, closed strings describe gravity in the 
bulk space-time, while open strings describe the D-brane dynamics. It is worth 
mention that at quantum level both strings are deeply connected.\\

It is then natural to study the role of closed string tachyons on the evolution 
of the Universe. Two interesting questions that we usually pursue in 
cosmology are: (i) Can such a closed tachyon drive the inflation or even the
acceleration of the Universe? and (ii) can this tachyon field be used to create a 
crunch scenario of the Universe? This is  expected as an analogy
with the open string tachyon
 \cite{Yang:2005rw,Yang:2005rx}; its condensation yields the 
disappearance of the source of tachyonic modes, i.e., D-branes. Therefore, 
it is proposed that the condensation of the closed string tachyon 
drives the disappearance of the source of tachyonic modes. In this case the 
object which perturbations are described by closed strings is the space-time 
itself. Many interesting studies on this topic have been performed in the last few 
years (see for instance \cite{McGreevy:2005ci,Swanson:2008dt, Aref'eva:2008gj}) on 
which the classical closed string tachyon field plays an important role in the evolution 
of the Universe.\\

The goal of this paper is to construct a cosmological toy model as an answer
for question (i). This model is based on the low energy limit of critical bosonic string 
theory, compactified on a non-flat manifold while keeping the dilaton constant. We present an 
exact solution of the classical equations of motion which contains 
interesting cosmological features.\\

Among other details, we find that cosmological issues depend on two parameters $A$ and $B$.  The tachyon potential (and in consequence the related cosmology), is partially determined by the curvature of the internal manifold which is proportional to $A^2$ and partially by the ratio $-A/B$ .\\

For a negative curved internal manifold and $-A/B\gg 1$,  the solution describes an accelerated  Universe which expands to a maximum size before collapsing. The initial and final singularities are reached in the infinite-past and infinite-future, respectively.
Inflation is guaranteed, while the Equation of State 
(EoS) for the tachyon is computable, showing that the closed tachyon field 
behaves as a cosmological constant at early times, changing its behavior 
as it approaches the minimum of the potential. \\ 

We explicitly show that the universe starts its contraction as the tachyon reaches the minimum of the potential, relating tachyon condensation with the collapsing of the space-time.
The condition, $-A/B\gg 1$ can be  accomplished by considering a compactification on an internal volume with a high curvature compared with the velocity at which the tachyon field evolves into the minimum of the potential.\\

We also mention briefly on the consistency of our solution in the context of quantum cosmology, since some features of our model, specifically the parameters on the tachyon potential, determine whether or not inflation is a generic feature in the evolution of the Universe. The subset of possible values of the parameters, determines a subset of the most general solutions, indicating that initial conditions play an important role in selecting such values. However, since at early times quantum gravity aspects must be considered, the question whether the obtained solution can be determined by initial conditions should be addressed in the context of quantum cosmology.\\

The outline of this paper is as follows: In Section \ref{closestring} we 
construct the effective action of a closed string tachyon and a gravitational 
field from bosonic string theory. Here we assume a constant dilaton and a 
constant internal volume. In Section \ref{tachyonversesec} we give the exact
solutions for the classical equations of motion of the scale factor and the 
tachyon field. We consider that the non-compact space-time is described by a 
spatially flat Friedmann-Lema\^itre-Robertson-Walker (FRLW) metric. In Section 
\ref{GEvo} we study the cosmological implications of our solutions, emphasizing the role played by the tachyon in the acceleration and inflation of the Universe, as questioned by point (i). We also study the EoS for the closed tachyon field and analyze an interesting 
example, in connection with the question (ii), in which the Universe contracts towards a big crunch.
Section V is devoted to address some quantum aspects of our solutions in the context of quantum cosmology.
 Section \ref{noconstvol} is devoted to comment on the case in which 
the internal volume is not constant and the future perspective on this matter. 
Finally, in section VII,  we give our conclusions and comments about future possible lines of study.
 

\section{Closed string tachyon}\label{closestring}

We start by considering  critical bosonic string theory in a $26$-dimensional 
space-time with a constant dilaton $\Phi_0$ and without dynamics related to a constant $B$-field. The 
action which describes the low-energy dynamics of the metric and the closed 
string tachyon field $T$ is therefore  given by 
\begin{equation}
S=\frac{1}{2\kappa_0^2}\int~d^{26}x\sqrt{-g_{26}}~e^{-2\Phi_0}\left[R-(\nabla T)^2
-2V(T)\right],
\end{equation}
where  $V(T)$ is the corresponding closed string tachyon 
potential and $g_{26}$ is the 26-dimensional metric \cite{Yang:2005rw, Yang:2005rx}.\\

By compactification on a non-flat internal 22-dimensional space $X_{22}$,  with constant curvature $\mathscr{R}$
the effective four-dimensional action reads
\begin{eqnarray}
S&=&\frac{1}{2\kappa_0^2}\int d^4x \sqrt{-g} ~Vol(X_{22})e^{-2\Phi_0}\times\nonumber\\
&&\left[R_4+\mathscr{R}-(\nabla T)^2-2V(T)\right],
\end{eqnarray}
where $Vol(X_{22})$ is the volume of $X_{22}$, $\kappa_0^2=8\pi G_{26}$ is the gravitational strength in 26 dimensions and the 26d metric is of the form
\begin{equation}
ds^2=g_{\mu\nu}dx^\mu dx^\nu+h_{mn}dy^mdy^n,
\end{equation}
with indices $\mu,\nu$ running from 0 to 3 and $m,n$ from 4 to 25.
Here we face two possible scenarios, concerning whether the internal compact 
space has a constant volume or not.  For the non-constant  case, it would be very difficult to exactly solve the field equations of motion, even for very simple choices of the tachyon potential. Henceforth, we shall consider 
the constant volume scenario without concerning the details on which the volume modulus is stabilized, 
although some  comments  are given in the last section.\\

Therefore, in a constant internal volume scenario, the effective 
four-dimensional action is
\begin{equation}
S=\frac{m_p^2}{2}\int d^4x~\sqrt{-g^E}\left[R_4\left(g^E\right)-(\nabla T)^2-2\mathscr{V}(T)\right],
\label{general action}
\end{equation}
where we have defined the reduced 4d mass Planck as
\begin{equation}
m_p^2=e^{-2\Phi_0} \frac{Vol(X_{22}) }{\kappa_0^2},
\end{equation}
and the effective scalar potential $\mathscr{V}$ is consequently given by
\begin{equation}
\mathscr{V}(T)= V(T) -\frac{1}{2}\mathscr{R}.
\end{equation}

Let us now choose the tachyon field to be a function only on time, i.e. $T=T(t)$, in 
order to guarantee the absence of $\alpha '$ corrections \cite{Swanson:2008dt}. Also we 
consider a spatially flat $3+1$ dimensional FLRW background
\begin{equation}\label{FRW metric}
\begin{aligned}
ds^{2}=-dt^{2}+e^{2\alpha(t)}\left(dr^{2}+r^{2}d{\Omega}^{2}\right),
\end{aligned}
\end{equation} 
where $a(t)=e^{\alpha(t)}$ is the scale factor. Substituting this metric in 
(\ref{general action}) the effective four-dimensional action can be written as
\begin{equation}\label{einstein3}
\begin{aligned}
S=m_p^2\int{\left[-3{\dot{\alpha}}^{2}(t) +\frac{1}{2}{\dot{T}}^{2}-\mathscr{V}(T)\right]e^{3\alpha(t)}dt}.
\end{aligned}
\end{equation}
The Einstein equations from action (\ref{einstein3}) and by using the FLRW metric 
(\ref{FRW metric}) are
\begin{eqnarray}\label{einstein-tach}
3\dot\alpha^2-\frac{1}{2}\dot{T}^2-\mathscr{V}(T)&=&0,\label{einstein-tach} \\
-2\ddot{\alpha}-3\dot\alpha^2-\frac{1}{2}\dot{T}^2+\mathscr{V}(T)&=&0,\label{einstein-tach2}
\end{eqnarray}
from which it is observed that $2\ddot\alpha+\dot{T}^2=0$, indicating the existence of 
a big crunch as the tachyon runs into larger values, addressing question (ii) in the Introduction. From action (\ref{einstein3}), the corresponding 
tachyon field equation of motion is
\begin{equation}
\ddot{T}+3\dot\alpha\dot{T}+\mathscr{V}'(T)=0.
\label{teom}
\end{equation}
%


\section{Classical closed string tachyon}\label{tachyonversesec}

In this section we discuss the canonical Hamiltonian formalism for  action
(\ref{einstein3}). Hereafter, we shall take the reduced Plank mass $m_p$ to be the unit. Then, with the usual definition for the momenta
\begin{eqnarray}\label{momenta}
	\pi_a &=& \frac{\partial L}{\partial\dot{\alpha}}= -6\dot{\alpha}e^{3\alpha}, \label{momenta1}\\
	\pi_T &=& \frac{\partial L}{\partial\dot{T}}= \dot{T}e^{3\alpha} \, \label{momenta2},
\end{eqnarray}
we are able to compute the corresponding canonical Hamiltonian:
\begin{align}
{\cal H}=\frac{e^{-3\alpha}}{12}\left[-\pi^2_\alpha+6\pi^2_T+12e^{6\alpha}\mathscr{V}(T)\right].
\end{align}
Using the standard Hamilton-Jacobi equation
\begin{equation}
H(\alpha, T; t) + \frac{\partial{S}}{\partial{t}}=0,
\label{WDW1}
\end{equation}
and by denoting the Hamilton's principal function as $S$, we make use of the standard 
identifications
\begin{equation}
\frac{\partial S(\alpha)}{\partial\alpha}=\pi_\alpha, \quad \frac{\partial S(T)}{\partial T}=\pi_T,
\label{pi}
\end{equation}
where $\pi_\alpha$ and $\pi_T$ are the canonical momenta given by (\ref{momenta1})
and (\ref{momenta2}). Therefore, in this approximation, equation (\ref{WDW1}) becomes,
\begin{equation}
-\frac{1}{12}(\partial_\alpha S)^2+\frac{1}{2}(\partial_T S)^2+e^{6\alpha}\mathscr{V}(T)=0,
\label{HJ}
\end{equation}
for $S$ not explicitly depending on time.
Now, we propose the function $S$ to be of the form \cite{Guzman:2005xt}
\begin{equation}
S(\alpha, T)= e^{3\alpha}W(T),
\label{S}
\end{equation}
where $W(T)$ is a polynomial function on $T$. There is a large arbitrariness in 
the choice of the potential for the closed string tachyon. Although there are
alternative proposals \cite{Yang:2005rw,Yang:2005rx}, specific tachyon 
potentials containing polynomial and trigonometric functions have been 
favored. All these models turn out to be very interesting, in that they 
admit a characteristic idea of cosmological evolution depending on the 
choice of initial conditions. Thus, in these models it was shown that the 
Universe evolves towards a big crunch singularity as deduced from Einstein 
equations (\ref{einstein-tach}) and (\ref{einstein-tach2}).\\

Here, we shall look for functions $W(T)$  such that, by substituting back into 
Eq.(\ref{HJ}), a string tachyon potential is obtained. Specifically we look for potentials which are polynomials on $T$ .  The corresponding differential equation is then given by
\begin{equation}
-\frac{1}{2}(\partial_T W)^2+\frac{3}{4}W^2=\mathscr{V}(T).
\label{potential}
\end{equation}
Generically, a polynomial potential is generated (via the above equation) by a function of the form
\begin{equation}
W(T)=(A+BT^m)T^n,
\label{W}
\end{equation}
for different choices of $n$ and $m$. 
However, the $T$-polynomial $W(T)$ must also be consistent with  the 
tachyon equation of motion (\ref{teom}). This restricts the set of possible solutions. We find that the simplest form of $W(T)$ satisfying the tachyon equation of motion is given by
\begin{equation}
W(T)=A+BT^2,
\end{equation}
for generic values of $A$ and $B$.

\section{Closed String Tachyon Cosmology}\label{GEvo}
We now wish to address a more realistic landscape with the above  proposal for $W(T)$.  The effective tachyon potential computed from Eq. (\ref{potential}) is given by
\begin{equation}
\mathscr{V}(T)= \frac{3}{4}(A+BT^2)^2-2B^2T^2.
\label{VT}
\end{equation}
This is quite interesting. First, notice that it is possible to relate the constant $A$ with the term $-\mathscr{R}$, indicating that the internal manifold has a negative curvature given by $\mathscr{R}=-3A^2/2$.  Eventhough we are preparing a simple toy model, it is probably of interest to remark that a negative curved internal manifold  has been shown to be an essential ingredient to obtain DeSitter classical vacua   in superstring flux compactifications \cite{Haque:2008jz, Danielsson:2009ff, Danielsson:2011au, Shiu:2011zt}. It is important  to stress out that in this model, the tachyon potential (and in consequence the related cosmology), is partially determined by the curvature of the internal manifold.  Therefore, the evolution of the Universe in this scenario, would be affected by the curvature of the extra dimensions.\\

Second, we see that this fact, $-$a negative internal curvature$-$, implies that the potential associated to the tachyon is precisely  the one expected from the string field theory perspective, i.e., a potential of the type $V(T)\sim -c_1^2T^2+c_2^2T^4$, for real numbers $c_1$ and $c_2$ as expected from string field theory \cite{Yang:2005rx}, which  vanishes at $T=0$. Finally, notice that positive curved internal spaces are not solutions of the classical equations of motion if the tachyon potential vanishes at $T=0$.\\

Some other interesting features of this potential concern the following: \\

\noindent
i) By substituting 
the function $W(T)$  back  into  (\ref{S}), we compute the canonical momentum $\pi_T$  from (\ref{pi}) and we solve 
the canonical momentum equation   for the tachyon (\ref{momenta2}). Therefore  the tachyon 
field is given by
\begin{equation}
T(t)= e^{2Bt},
\label{Tt}
\end{equation}
where we have fixed the integration constant to one. Using this expression, it is 
straightforward to compute the scale factor $a(t)=e^{\alpha(t)}$ from the 
equation of motion for $\alpha(t)$ which reads
\begin{equation}
a(t)= e^{\left(-\frac{1}{2}At-\frac{1}{8}e^{4Bt}\right)}.
\label{scaleft}
\end{equation}

\noindent
ii) The effective potential $\mathscr{V}(T)$ has a minimum at $T_0=\sqrt{(4B-3A)/3B}$ . Therefore, a solution involving $A>0$ and $B<0$ is excluded. The value of the potential at the minimum is 
\begin{equation}
\mathscr{V}(T_0)=\frac{2}{3}B(3A-2B),
\end{equation}
from which it is observed that  the vacuum energy is negative for $A<0$ and $B>0$. Otherwise, the vacuum energy can be positive, null or negative. The time at which the tachyon field reaches the minimum of the potential is
\begin{equation}
t_T=\frac{1}{4B}\ln{\left(-\frac{A}{B}+\frac{4}{3}\right)}.
\end{equation}

\noindent
iii) We see that an internal flat 22d space (i.e., $A=0$) is an available solution. The tachyonic potential is given in this case by 
\begin{equation}
\mathscr{V}(T)|_{A=0}=-2B^2T^2+\frac{3}{4}B^2T^4,
\end{equation}
with $B>0$. This fact supports our previous identification of the scalar factor $3A^2/2$ in the potential (\ref{VT}) as the internal curvature. Moreover, for this case,  at $t\rightarrow -\infty$  the Universe has a size given by $a(-\infty)=1$.  Therefore, in these solutions the Universe  starts without the presence of a singularity, avoiding a Big Bang scenario.
\\

\noindent
iv) The time $t_0$ at which the scale factor reaches a maximum, is given by
\begin{equation}
t_0= \frac{1}{4B}\ln{\left(-\frac{A}{B}\right)},
\end{equation}
from which we conclude that unless $A/B$ is negative, the Universe will not face a change from its initial contraction or expansion stage.   Therefore, a change of stage is accomplished by taking $A<0$ and $B>0$. In such case, the scale factor describes a Universe which expands from an asymptotically zero-volume (i.e., $a=0$ at $t\rightarrow -\infty$), reaches a maximum size at the time  $t_0$ and contracts to a point in an infinite time. We shall consider this case hereafter. \\

\noindent
v) The parameter $A^2$ in the effective tachyon potential $\mathscr{V}(T)$ plays the role of a cosmological 
constant for very early times ($t \rightarrow -\infty$). However, its value is 
not directly related to the vacuum energy, since this latter can be positive 
even if $A$ is negative as mentioned above. \\

Our next step is to study whether the collapse of the Universe is related to the tachyon condensation and the conditions on which acceleration and inflation are present in these scenarios.


\subsection{Acceleration and Inflation}

Specific cosmological issues are dependent on the values of the parameters 
$A$ and $B$. An acceleration stage is present if the ratio $\ddot{a}/a$ is positive, which in accordance with Eqs.(\ref{Tt}) and (\ref{scaleft}) is given by
\begin{equation}
\frac{\ddot{a}}{a}=\frac{1}{4}[B^2e^{8Bt}+2B(A-4B)e^{4Bt}+A^2].
\end{equation}
The ratio  is positive for $t\rightarrow -\infty$ till $t=t_a$ with
\begin{equation}
t_a= \frac{1}{4B}\ln \left(\frac{4B-A-2\sqrt{4B^2-2AB}}{B}\right),
\end{equation} 
time at which  the acceleration stage finishes. Remember we are considering the case in which $A<0$ and $B>0$. \\

Interesting to notice is that for generic values of the parameters $A$ and $B$, we have that $t_a<t_0<t_T$, indicating that the Universe reaches its maximum before the tachyon condensates. The condensation occurs after the end of the acceleration stage.   However, for $-A/B\gg 1$, $t_a\approx t_0\approx t_T$. Therefore in such scenarios, the end of an accelerated Universe, the tachyon condensation and the beginning of the collapsing, all events take place  almost simultaneously.\\

The question is now if there exists conditions for an inflation period within this accelerated stage of the Universe. Particularly we would like to search for inflationary slow-roll conditions driven by the effective tachyon potential $\mathscr{V}(T)$.
Necessary and sufficient conditions for 
inflation are given by   $\epsilon\ll1$ and  $\eta\ll1$ 
where
\begin{eqnarray}
 \epsilon(T)
 &=&\left(\frac{1}{\mathscr{V}}\frac{\partial \mathscr{V}}{\partial T}\right)^2 ,\nonumber\\
 &=&\frac{16B^2T^2(3A-4B+3BT^2)^2}{[3A^2+2BT^2(3A-4B)+3B^2T^4]^2 },
 \end{eqnarray}
and 
\begin{eqnarray}
|\eta(T)|&=&\left|\frac{1}{\mathscr{V}}\frac{\partial^2 \mathscr{V}}{\partial T^2}\right|,\nonumber\\
&=&\frac{|(3A-4B)B+9B^2T^2|}{|\frac{3}{4}(A+BT^2)^2-2B^2T^2|},
\end{eqnarray}
for $\mathscr{V}>0$. Inflation is also assured if  $\dot{\mathscr{V}}\ll 1$. It turns out that 
there are several values of $A$ and $B$ for which such conditions are assured 
during a specific time interval.\\

For generic values of $A$ and $B$, there are not always conditions on which 
inflation can be carried out. However, we can find some ranges on values of $A$ and $B$ on which inflation is guaranteed.
 This is specified by the  ratio $A/B$ which not only  determines how fast the Universe expands 
and/or collapses but also indicates the presence or absence 
of inflation. It turns out that for $|A|/B\geq 1$ the slow-roll conditions are fulfilled. The details are shown in the appendix.


\subsection{Insight into the Equation of State}\label{EoStachyonverse}

In order to obtain cosmological implications, we assume that the closed 
tachyon is a scalar field evolving in a FLRW Universe which contains a fluid 
described by the Equations of State (EoS) $p=\omega\rho$, where $\rho$ is the energy density, 
$p$ the pressure and $\omega$ is a constant such that  a dust ($\omega=0$), 
radiation ($\omega=1/3$) or a cosmological constant ($\omega=-1$) dominated universe can be described. Therefore, 
the Hubble term in this scenario is
\begin{equation}\label{hubble2}
\begin{aligned}
	H=\frac{\dot{a}}{a}=\dot\alpha=-\frac{1}{2}\left(A+Be^{4Bt}\right),
	\end{aligned}
\end{equation}
which for very early times ($t\rightarrow -\infty$) behaves as a positive constant. On 
the other hand, the expressions for the pressure and density are computed by 
the energy-momentum tensor $T_{\mu\nu}$ are
\begin{eqnarray}
	P_{T} &=& \frac{1}{2}{\dot{T}}^{2}-\mathscr{V}(T) \, \label{pressure} ,  \\
	\rho_{T} &=& \frac{1}{2}{\dot{T}}^{2}+\mathscr{V}(T) \, \label{density} , 
\end{eqnarray}
where the subindex $T$ denotes the relation with the closed string tachyon. 
Replacing the potential (\ref{VT}) and the solution (\ref{Tt}) in (\ref{pressure}) 
and (\ref{density}) we obtain the following expressions for the pressure, 
density and EoS of the closed string tachyon, respectively,
\begin{eqnarray}
P_{T} &=& 4B^2e^{4Bt}-\frac{3}{4}(A+Be^{4Bt})^2 \, \label{pressure2} , \\
\rho_{T} &=& \frac{3}{4}(A+Be^{4Bt})^2 \, \label{density2} , \\
\omega_{T} &=&-\frac{3A^2+2Be^{4Bt}(3A-8B)+3B^2e^{8Bt}}{3(A+Be^{4Bt})^2} \, \label{equationstate} .
\end{eqnarray}
We observe that at $t\rightarrow -\infty$,  with  the tachyon field  at 
the top of the potential, $\omega_T=-1$, indicating that at early times  the tachyon potential
 behaves as a cosmological constant. This also follows from the tachyon 
field potential $\mathscr{V}(T)$ from which we see that at early times, the only term which 
survives in the action is the constant $A^2$ which precisely behaves as a cosmological 
constant.\\

Let us comment on  the complete evolution of (\ref{pressure2}), (\ref{density2}) and 
(\ref{equationstate}) together. At the beginning, the density $\rho_{T}$ is 
positive, $\rho(t\rightarrow -\infty)=3A^2/4$. As time runs, $\rho$ decreases 
indicating an expansion stage. The minimum value of $\rho$ is zero and it is reached at 
$t=t_\rho$ where
\begin{equation}
t_\rho= \frac{1}{4B}\ln{\left(-\frac{A}{B}\right)},
\end{equation}
which is precisely the time $t_0$ at which the Universe reaches its maximum size. \\

Notice also that the pressure $P_{T}$ is negative at $t\rightarrow -\infty$ and 
increases its value as the Universe evolves until a maximum positive value at 
time $t=t_{p}$. This is consistent with a Universe which expands to a 
maximum size. After that, i.e. for times greater than $t_p$, the pressure 
decreases, and the Universe starts a contraction stage, becoming more and more 
negative as the time evolves, until the space-time itself blows up 
\cite{Yang:2005rw}. The time at which the pressure is maximum is given by
\begin{equation}
t_p=\frac{1}{4B}\ln{\left(\frac{8}{3}-\frac{A}{B}\right)}.
\end{equation}
Hence, in the generic case, we have that $t_a<t_0=t_\rho<t_T<t_p$. However, 
as mentioned previously, for $-A/B\gg 1$ it happens that $t_p\approx t_\rho\approx t_T\approx t_a\approx t_0$, meaning that the time at which the collapse of the Universe starts, defined by the time at which the pressure and density vanishes, coincides with the time at which the tachyon field reaches its minimum in the tachyonic potential, i.e., when it condensates. \\

Such condition, $-A/B\gg 1$ can be  accomplished by considering a compactification on an internal volume with a high curvature compared with the velocity at which the tachyon field evolves into the minimum of the potential.\\

Important to notice, is that inflationary conditions depend solely on the values of the tachyon potential parameters $A$ and $B$ and do not on the values of $\dot{T}$, which is always very small at early times. \\

\subsection{An example}

\begin{figure}[t]
\begin{center}
\centering \epsfysize=8cm \leavevmode
\epsfbox{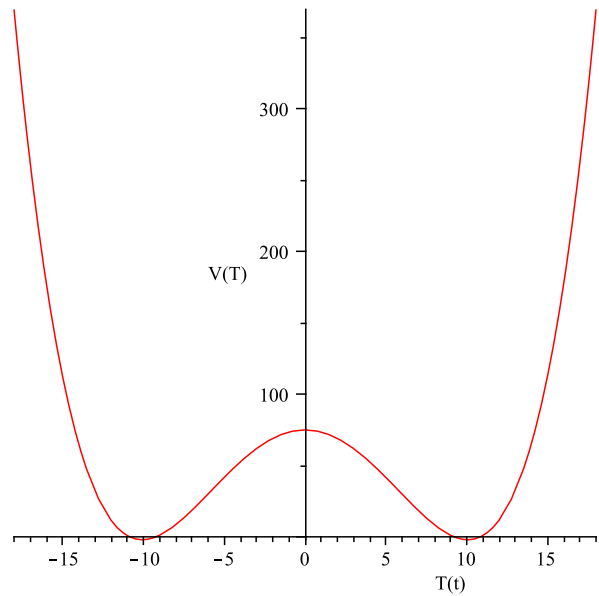}
\end{center}
\caption[hola]{\small \it Tachyon field potential for $A=-10$ and $B=1/10$.}
\label{t1}
\end{figure}

\begin{figure}[t]
\begin{center}
\centering \epsfysize=8cm \leavevmode
\epsfbox{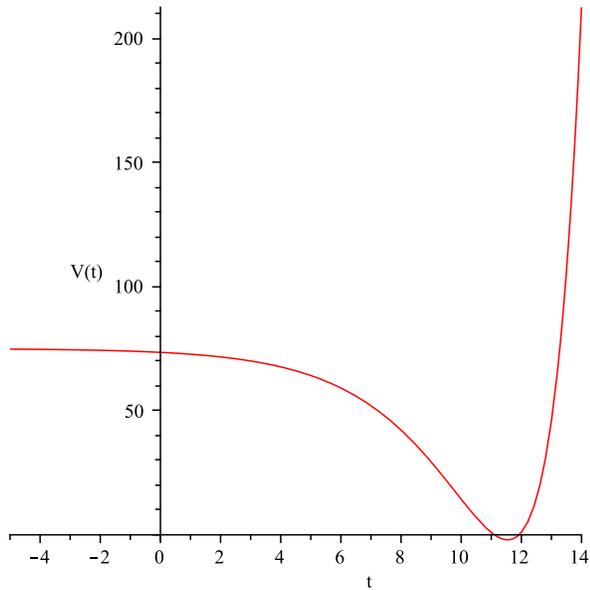}
\end{center}
\caption[hola]{\small \it Tachyon potential as a function of time.}
\label{t2}
\end{figure}

Consider the following example: $A=-10, B=1/10$. The tachyon potential depicted 
in Figure (\ref{t1}), is given by
\begin{equation}
\mathscr{V}(T)= A_0T^4+B_0T^2+C_0,
\end{equation}
where $A_0=3/400, B_0=-38/25$ and $C_0=75$. From the corresponding equations 
of motion, we obtain that $T(t)=e^{t/5}$, yielding the tachyon potential in 
terms of $t$ as
\begin{equation}
\mathscr{V}(t)= C_0-B_0e^{2t/5}+A_0e^{4t/5},
\end{equation}
as is shown in Figure (\ref{t2}).  The tachyon field reaches the minimum of the 
potential at $t= 5\ln(304/3)^{1/2}$.\\

\begin{figure}[t]
\begin{center}
\centering \epsfysize=8cm \leavevmode
\epsfbox{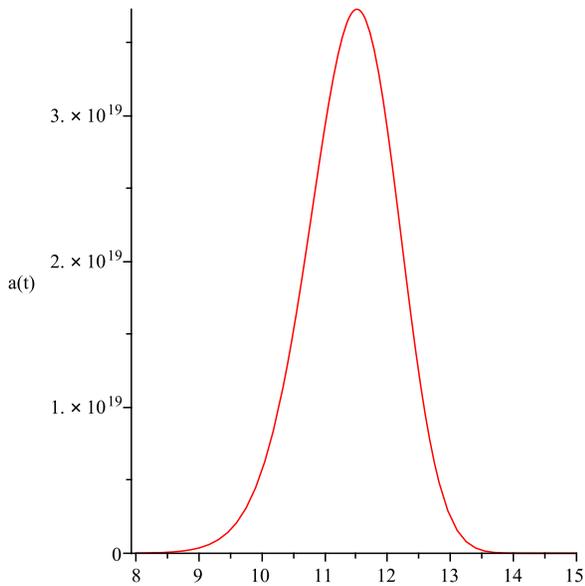}
\end{center}
\caption[hola]{\small \it Scale factor as a function of time.}
\label{scalef1}
\end{figure}

Under these conditions, the scale factor of the Universe is given by 
\begin{equation}
a(t)= e^{\left(5t-\frac{1}{8}e^{\frac{2t}{5}}\right)},
\end{equation}
which describes a Universe which expands for some time till $t_\ast=5\ln(10)$ 
after which a contraction towards a big crunch starts, as in Figure (\ref{scalef1}).\\

\begin{figure}[t]
\begin{center}
\centering \epsfysize=8cm \leavevmode
\epsfbox{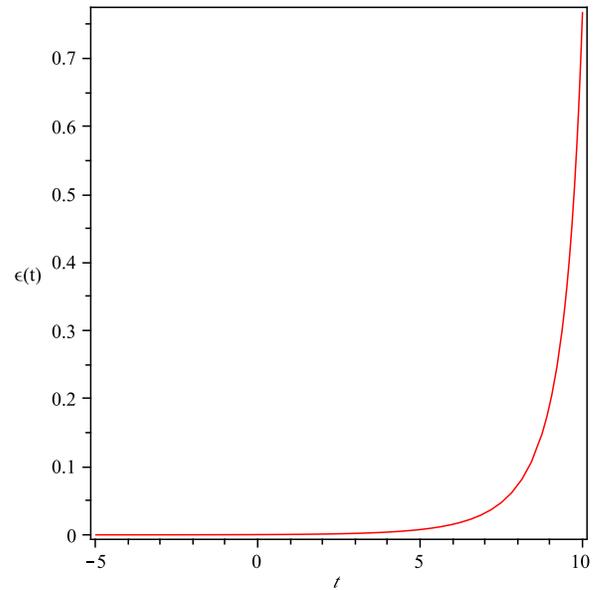}
\end{center}
\caption[hola]{\small \it Parameter $\epsilon$ as a function of time.}
\label{epsilon}
\end{figure}

\begin{figure}[t]
\begin{center}
\centering \epsfysize=8cm \leavevmode
\epsfbox{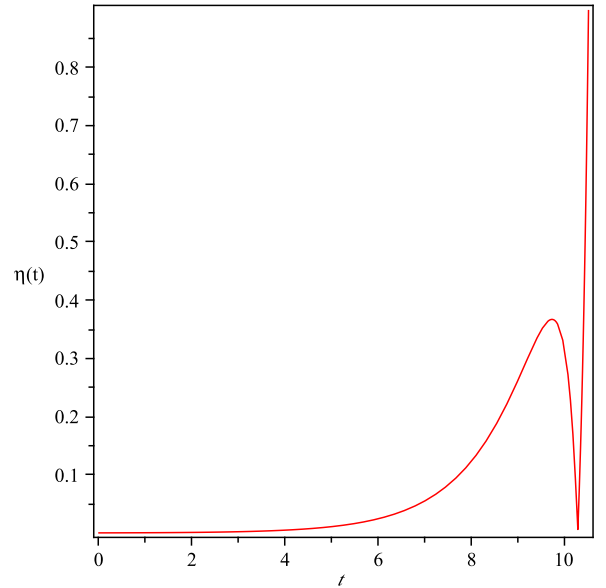}
\end{center}
\caption[hola]{\small \it Parameter $\eta$ as function of time.}
\label{eta}
\end{figure}

We can observe that under the period of expansion, the related potential looks 
very flat, indicating that inflation conditions are suitable to be found. Indeed, 
the slow-roll conditions are present for times $t\in(-\infty, 10)$. This interval 
guarantees small values for $\eta$ and $\epsilon$. In this way, we see that this model presents 
at the beginning a cosmological inflation, predicting as well the time at which 
it finishes. The evolution of the values for $\epsilon$ and $\eta$ are 
shown in Figures (\ref{epsilon}) and (\ref{eta}).\\

Notice as well that in those intervals, the scalar potential as a function of time $\mathscr{V}(t)$ is positive and large, and 
$\partial \mathscr{V}/\partial t <1$. Important enough, is to notice that just before 
the collapses starts, the energy of the Universe is positive. Therefore, 
tachyon condensation leads to a Universe which expands on a certain period 
of time $-$within an inflationary stage is present$-$ before it starts 
collapsing. Some time before the tachyon reaches an stable extreme at the 
potential, the Universe starts its collapse, and the value of the vacuum energy is 
very small and positive. As time runs forward, the vacuum energy turns smaller. At the end, i.e., at the minimum of the potential, the vacuum energy is negative but very close to zero. Actually is closer to zero as the quotient $A/B$.\\

\begin{figure}[t]
\begin{center}
\centering \epsfysize=8cm \leavevmode
\epsfbox{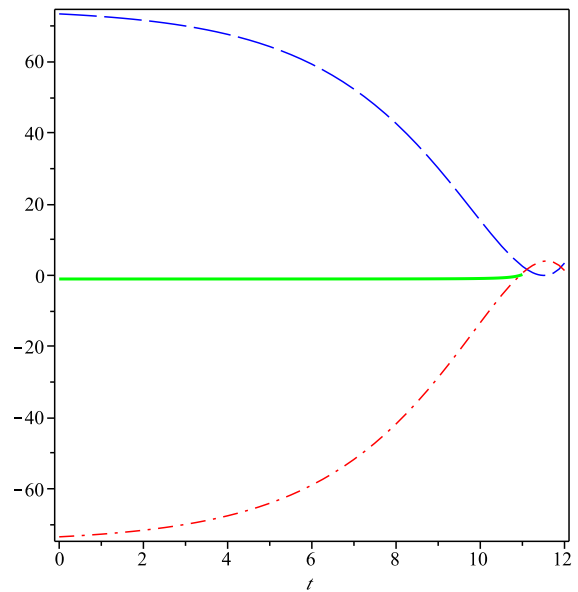}
\end{center}
\caption[hola]{\small \it {\it Solid green line}: the factor $\omega$ which for 
very early times is equal to -1. {\it Dash blue line}: energy density and 
{\it dash-point red line}: the pressure.}
\label{eos}
\end{figure}

Finally, it is possible to analyze the energy density, pressure and the 
corresponding EoS for this model, which are given by
\begin{eqnarray}
\rho &=& 3\left[5-\frac{1}{20}e^{\left(\frac{2t}{5}\right)}\right]^2,\\
p&=&\frac{1}{25}e^{\left(\frac{2t}{5}\right)}-3\left[5-\frac{1}{20}e^{\left(\frac{2t}{5}\right)}\right]^2,\\
\omega&=&\frac{616 e^{\left(\frac{2t}{5}\right)}-30000 -3e^{\left(\frac{4t}{5}\right)}}{3\left[-100+e^{\left(\frac{2t}{5}\right)}\right]^2}.
\end{eqnarray}
The corresponding behavior is shown in Figure (\ref{eos}). As mentioned in the 
general case, this tells us that the tachyon field behaves as a cosmological constant for 
very early times until the tachyon condensates at
\begin{equation}
t_\ast\approx5\ln (10),
\end{equation}
 time at which the Universe 
starts its collapse. Also we see the typical behavior of a Universe expanding 
and collapsing. The energy density decreases as the Universe expands and 
increases as soon as the tachyon condensates, indicating a collapse. The 
pressure is negative before the tachyon condensation, describing a Universe 
which expands, but its (absolute) value becomes smaller until it reaches a time in which 
there is no pressure to become then positive. This describes a stage in which the 
Universe expands, it slows down before stoping and turning into collapse. \\

\section{Quantum Cosmology}
We have seen that the exact classical solution of the Hamilton-Jacobi equation yields a Universe with a dynamics
driven by an asymptotically vanishing tachyon field at $t\rightarrow-\infty$. We have also found that for some subset of values of the parameters $A$ and $B$, inflation conditions are assured.  At very early times, one expects that initial conditions would determine the values of  the parameters $A$ and $B$ in the tachyon potential and in consequence the presence or absence of inflation.\\

However, at very early times, quantum gravitational effects become important, leading us to rethink on the validity of our solutions. For instance there is not certainty that our classical exact solution is selected by quantum initial conditions, or that this classical solution is  compatible at quantum level, meaning that, since our solution is exact, we expect that it must be approximately valid for all regions, even at those on which quantum effects become important.\\

All these features must be studied in a framework which considers quantum effects, as in quantum cosmology models (see \cite{Halliwell:2009rb} and references therein).  In this context,  the system is quantized by the introduction of a wave function $\Psi$ associated to a Universe as a solution of the Wheeler-DeWitt (WDW) equation $\hat{H}|\Psi\!\!\!>=0$, which is the equation relevant to quantum  cosmology and represents the quantum version of the reparametrization invariance in the classical theory.\\

 There are multiple solutions for this equation, corresponding to different initial conditions. By fixing such conditions, the hope is that a unique solution to the WDW equation would be selected. This however has turn very difficult to prove. Nevertheless,  there exist in the literature two different proposals which attempt to develop a method  to choose a particular solution of the WDW equation \cite{Hartle:1983ai, Vilenkin:1987kf}. In both cases, the quantum description of initial conditions must imply the existence of a universe which is approximately classical when it is large.\\


One of such frameworks, which for simplicity we sutdy here,  is given by the so called {\it tunneling boundary condition} which involves the study of the behavior of the solutions of the WDW equation at the boundary of the superspace. Roughly speaking the proposal attempts to classify the solutions of the WDW equation as {\it ingoing} or {\it outgoing} at the boundary of the superspace together with a condition on the wave function $\Psi$ to be everywhere bounded. Those solutions which fulfill these requirements can be considered a well-behaved wave function with appropriate initial boundary conditions   \cite{Halliwell:2009rb}.

\subsection{WKB approach}
In order to check that our exact classical solution is indeed compatible with the framework of quantum cosmology, the first step consist on reproducing such solution as an approximate solution to the WDW equation.  In consequence, we must focus on a region where a classical description is expected to be valid. This corresponds to large  values of the scale factor $a(t)$, or equivalently, to times far away from initial and final stages of the universe.\\

Let us start by promoting the canonical momenta to differential operators in the WDW equation,
\begin{equation}
\frac{e^{-3\alpha}}{12}\left[-\hat\pi^2_\alpha+6\hat\pi^2_T+12e^{6\alpha}\mathscr{V}(T)\right]\Psi=0,
\end{equation}
and look for solutions of the form $\Psi=e^{iS}$, with $S=S(\alpha, T)$.\\

Inserting this into the WDW equation, and by taking the WKB approximation, one finds that $S$ obeys the Hamilton-Jacobi equation, in particular with $S=W(T)e^{3\alpha}$. This tells us that our classical solution is indeed a solution for the WDW equation in the approximation  $\Psi=e^{iS}$.  Observe that the WKB approximation implies that
\begin{equation}
BT^2e^{3\alpha}\gg 1,
\end{equation}
which holds at the classical region as expected, independently of the value of $B$.\\

Conceptually, this means that among all possible solutions of the WDW equation, there is a subset of solutions about which this solution is peaked. In turn, the selection of such solution implies the existence of quantum initial conditions. In pedestrian words, quantizing the system leads us to transfer the problem of initial conditions from a classical perspective  to a quantum description.\\

Therefore, the approximate solution of the WDW equation is given by
\begin{equation}
\Psi=e^{[i(A+BT^2)e^{3\alpha}]}.
\end{equation}
This would represent a good solution of quantum cosmology,  according with the tunneling boundary conditions, i.e.,  if it represents an outgoing wave function from the boundary which is everywhere bounded.

\subsection{Tunneling boundary conditions}

Therefore, our second step is to prove that $\Psi$ consists solely of outgoing modes at singular boundaries of superspace. For that we shall show that the wave function is oscillatory in the neighbourhood of the boundary, which makes easier to define an outgoing mode.\\

For early times, it is easy to see that $\partial_T S$ is almost negligible. Therefore the WDW equation reduces to
\begin{equation}
\left[-(\partial_\alpha S)^2+12 e^{6\alpha}\mathscr{V}(T)\right]\Psi=0,
\end{equation}
where $\mathscr{V}(T)$ is always positive. In the WKB approximation, one solution of this equation is given by
\begin{equation}
\Psi\approx \exp\left(\pm\frac{2i}{\sqrt{3}}e^{3\alpha}\sqrt{\mathscr{V}(T)}\right).
\end{equation}
This oscillatory solution is the only one solution given that the potential is always positive. However only $\Psi=e^{-iS}$ represents an outgoing mode \cite{Halliwell:2009rb}. \\

Finally, we must verify if this approximated solution is regulated at the boundary of the superspace, this is, if $\partial \Psi/\partial T\rightarrow 0$ as $a\rightarrow 0$. Indeed, we see that
\begin{equation}
\frac{\partial \Psi}{\partial T} \approx -\frac{a}{\sqrt{3}}\frac{2BT}{\mathscr{V}(T)},
\end{equation}
which vanishes for $a\rightarrow 0$. This analysis confirms that our classical solution is indeed well behaved, since $\Psi$ is an oscillatory and regular wave function in a classical allowed region, implying that our exact solution is valid on regions on which quantum effects are important.\\




\section{Comments on a non-constant internal volume}\label{noconstvol}

As it was mentioned at the beginning, we have assumed that the internal manifold 
posses a constant volume. This assumption is not justified through a dynamical 
process which stabilize the internal volume. For that we have learn in the past 
years that turning on internal fluxes in superstring compactifications on manifolds with a generalized structure, is the 
easiest perturbative way to stabilize some moduli as the internal volume (K\"ahler moduli). Therefore 
we expect that in this case $-$a bosonic string compactification$-$ the NS-NS 
flux ${\mathcal{H}}_3$ would generate an effective potential 
which depends on the volume modulus. \\

The corresponding low energy action from critical bosonic string theory in the presence of a tachyon field and a NS-NS flux  $H_3$ is given by
\begin{equation}
\frac{1}{2\kappa_0^2}\int d^{26}x~\sqrt{-g}~e^{-2\Phi_0}\left[R-\frac{1}{12}H^2-(\nabla T)^2-2V(T)\right],
\label{Haction}
\end{equation}
where we have assumed a constant dilaton (see \cite{Suyama:2006wx} for a recent study).\\


In order to reproduce the 
usual 4d Hilbert-Einstein action after a compactification on a non-flat space $X_{22}$, we require to write the above action in the 
Einstein frame by a conformal transformation \cite{Faraoni:1998qx}.
The Weyl transformation is taken to be
\begin{equation}
g_{\mu\nu}=\Omega^2 g^E_{\mu\nu},
\end{equation}
where $g^E$ is the metric in the Einstein frame. By defining the effective 
internal volume $\tau$ by
\begin{equation}
\tau^2=e^{-2\Phi_0}\rho^{11},
\label{frame}
\end{equation}
with $\rho^{11}=Vol(X_{22})$, and by taking
\begin{equation}
\Omega^2=\frac{m_p^2\kappa_0^2}{\tau^2},
\end{equation}
where $m_p$ is the reduced Planck mass in four dimensions, the effective action becomes
\begin{eqnarray}
S&=&\frac{m_p^2}{2}\int d^4x~\sqrt{-g^E}\left[R_4(g^E)+6(\nabla \ln~\Omega)^2
+\right.\nonumber\\
&&\left.6\nabla^2\ln\Omega-(\nabla T)^2-2\mathscr{V}(\tau,\rho,T)\right],
\label{noconstant}
\end{eqnarray}
where $\mathscr{V}$ is the effective potential and it is a function on the moduli $(\tau,\rho, T)$ to be determined. Notice that for a non-constant dilaton $\Phi$, 
$\tau$ and $\rho$ would represent two independent moduli. However, in this case 
they are correlated, implying that
\begin{equation}
\partial_\mu \ln\tau=\frac{11}{2}\partial_\mu \ln\rho.
\end{equation}
Using this relationship, the effective scalar curvature in the Einstein frame 
reduces to
\begin{equation}
R_4(g)= R_4\left(g^E\right)+6\left(\nabla \ln\tau\right)^2
-6\nabla^2\ln\tau.
\end{equation}

In the same context, the effective potential $\mathscr{V}$  (in the Einstein frame)
depends only on two parameters: the tachyon field $T(t)$ and the internal modulus $\rho$, and it can be written as  \cite{Hertzberg:2007wc}
\begin{equation}
\mathscr{V}(T, \rho)= \frac{\mathcal{H}(\phi_0)}{\rho^{14}}+\frac{W(\phi_0)}{\rho^{11}}\left[V(T)-\frac{1}{2}\mathscr{R}\right],
\label{potH}
\end{equation}
where $\mathscr{R}$ is the internal curvature.  
The functions ${\cal H}(\phi_0)$ and $W(\phi_0)$ are given by
\begin{eqnarray}
{\cal H}(\phi_0)&=& \frac{m_p^2\kappa_{0}^2}{12} e^{2\Phi_0}|H^2|,\nonumber\\
W(\phi_0)&=&  m_p^2\kappa_{0}^2~e^{2\Phi_0}.
\end{eqnarray}
By considering a  tachyonic potential $V(T)$ of the form $V(T)=-c_1T^2+c_2T^4$,  with $c_1$ and $c_2$ being  positive numbers, some interesting issues  can be deduced from a general treatment. 
For instance, we see that in the absence of $H$-flux, the potential has not a minimum 
with respect to $\rho$. This enforces the expected result that 
volume is stabilized through the presence of $H$. Actually, the potential $\mathscr{V}(T,\rho)$ 
has a minimum at
\begin{eqnarray}
T_0&=& \sqrt{\frac{c_1}{2c_2}},\\
\rho_0&=&\frac{2(847)^{1/3}}{11}[({\cal H}c_2)(4c_2\mathscr{R}+c_1^2)]^{1/3}.
\end{eqnarray}
In order to have a positive internal volume, it is required that
$4c_2\mathscr{R}+c_1^2> 0$. Hence, the internal curvature can be positive, null or negative, contrary to the fluxless case we studied in the previous sections.  However, a full treatment of 
this case would require solving the corresponding differential equations obtained 
from  action (\ref{Haction}) by taking a non-constant internal volume.\\

Back to our case, the vacuum energy value is then given by 
\begin{equation}
\mathscr{V}(T_0, \rho_0)^3=-K_0 \frac{(c_1^2+4c_2\mathscr{R})^{14}}{{\cal H}^{11}c_2^{14}},
\end{equation}
which is always negative for any value of $c_1$, $c_2$ and $\mathscr{R}$, which in turn fulfill the positive-volume constraint $c_1^2+4c_2\mathscr{R}>0$. $K_0$ is a positive constant.\\

The question now is whether  the potential is positive just before
the tachyon field reaches its minimum at the tachyon potential. For that, let 
us assume that the modulus $\rho$ is stabilized at the value $\rho_0$ before the tachyon condensates. At such point, and assuming that the tachyon fields increases its value (with respect to a parameter as time) as it approaches the minimum , it can be written as
\begin{equation}
T_1=T_0-\kappa,
\end{equation}
for $\kappa\ll 1$. In this case, the value of the potential is (up to second 
order)
\begin{eqnarray}
\mathscr{V}= -K\frac{(c_1^2+4c_2\mathscr{R})^{11/3}(3c_1^2+12c_2\mathscr{R}-112c_1c_2\kappa^2)}{c_2^{14/3}{\cal H}^{11/3}},
\end{eqnarray}
with $K$ a positive number. Therefore,  just before the 
tachyon condensates,  the vacuum energy  is positive  for
\begin{equation}
 \kappa^2> \frac{3(c_1^2+4c_2\mathscr{R})}{112 c_1c_2}.
\end{equation}
For negative curvatures (for instance if  $\mathscr{R}$ is approximately $-c_1^2/4c_2$), it is easy to fulfill the condition $\kappa \ll 1$. In the case in which tachyon condensation is related 
to a Big Crunch, the above analysis would indicate the possibility to have a Universe with a 
tiny positive cosmological constant just before the collapse starts. A more 
detailed analysis of this case is under progress \cite{coo}.\\



\section{Conclusions}

In this paper we have considered a critical bosonic string compactification into a four-dimensional space-time by taking a constant dilaton, no extra fluxes and a non-flat internal manifold. Our main goal is to study the role played by the closed string tachyon field in the cosmology of the effective space-time.\\

By selecting a time-like tachyon field, in a background governed by a flat 3-dimensional FLRW space-time, we find exact solutions of the equations of motion of the scale factor $e^{\alpha}$ and the tachyon field $T$, both implicitly depending on time. For that,  we make use of the Hamilton-Jacobi (H-J) equations by selecting as the principal function $S$ a product of functions depending on $\alpha$ and $T$, Eq.(\ref{S}), explicitly given by $W(T)e^{3\alpha}$.\\

This form of the principal H-J function, establishes a differential equation by which the effective potential $\mathscr{V}$ is determined. We find that $\mathscr{V}$ consists on a negative constant, related to the internal curvature $\mathscr{R}$, and a closed string tachyonic potential of the form $V(T)\sim -a^2T^2+b^2T^4$, which is the expected closed string tachyonic potential vanishing at $T=0$.\\
 
 The corresponding solutions for the tachyon field and the scale factor depend on two parameters $A$ and $B$,  and they consist on   an exponentially growing tachyon field and a scale factor which behavior is determined by the curvature of the internal manifold $\mathscr{R}=-3A^2/B$ and the ratio  $-A/B$. \\

Interestingly, we find that for $A<0$, $B>0$ and $-A/B\gg 1$, there are conditions to have a universe which expands  to a maximum value before collapsing.  By parametrizing the solution with respect to time, we find also that the contraction stage begins at times  at which the tachyon reaches its minimum in its potential $V(T)$. This model enforces the  idea on which the closed string tachyon condensation is related to the collapse of the space-time.
Also we find that under these conditions, inflation is always present in the evolution of the universe since the standard slow-roll conditions are fulfilled within a period of time before the Universe collapses.\\

Finally we study the evolution of the Equation of State of the closed string tachyon by identifying the tachyon contributions to the Einstein equations as a perfect fluid. Under this assumption, it is found that at early times, the tachyon potential behaves as a cosmological constant, driven the universe to an expansion inflationary stage.
Therefore, we see from the EoS that the time at which the universe starts its collapse coincides with the time at which the tachyon reaches the minimum of the potential and it also coincides with the time at which the accelerations stops. \\

There we clearly see that positive tiny values for the cosmological constant are always possible at times just before the tachyon condensates if we compactify on manifolds with a high negative curvature. Only with negative curvature it is possible to obtain tiny cosmological values.  However, within this scenario, it is not possible to obtain positive values for the vacuum energy. In order to have a positive vacuum energy it would be necessary to consider quantum corrections. It is worth mentioning that we have assumed a flat 3-dimensional space metric as a background. That means that the universe described by this model is always flat.\\

On the other hand, it is interesting to observe that at early times, for arbitrary values of the parameters $A$ and $B$ , we have a tachyon field which evolves slowly, i.e., $\dot{T}\approx 0$. This turned to be important once we considered quantum aspects of our solutions. In this context, we find that the classical exact solution is recovered from the WKB approximation of the Wheeler-DeWitt equation in quantum cosmology. Moreover, we show that such solution turns out to be compatible with a quantum theory of initial conditions, implying that it is valid on all regions even on those on which quantum effects become important, at least in the context of quantum cosmology.\\

We also  comment at the end of our work about the case in which the internal manifold is not constant.  By a qualitative analysis we confirm that the existence of a tiny cosmological constant at the minimum of the potential is related to a compactification on a negative-curved space. The higher the absolute value of the curvature, the smallest value of the vacuum energy.\\

A lesson we take away is that the closed string tachyon provides many possible  
future studies if we consider a variety of interacting moduli cosmologies 
and if we keep in mind that there is not need for an assumption about the form of the 
tachyon potential, which is  totally fixed by the model itself once we consider a compactification on a negative curved manifold.


\begin{acknowledgments}

We thank M. Sabido and I. Quiros for very interesting observations and discussions. C. E.-R. would like to thank Pedro G. Ferreira for his kind 
hospitality in Oxford. C. E.-R. is supported by Fundaci\'on Pablo Garc\'ia, FUNDEC, 
M\'exico and partially by Basque Goverment AE-2010-1-31, O. L.-B. is 
partially supported by PROMEP under grant No. UGTO-PTC-207. O.O. is partially supported by CONACyT under grant No. 135023 and by PROMEP.

\end{acknowledgments}

\appendix
\section{Slow-roll conditions for inflation}
We look for generic conditions on the parameters $A$ and $B$ under which the slow-roll inflationary conditions are fulfilled. First, notice that $\epsilon$ is a positive even function on $T$. Hence, its minimum value is obtained at $T=0$. At $T=0$ the tachyon potential has a maximum and the parameter $\epsilon$ vanishes. For larger values of $t$ with $t\geq0$, the tachyon field $T(t)\geq 0$. One way to look for conditions on which $\epsilon\ll 1$ is to guarantee that $\partial \mathscr{V}/\partial T< \mathscr{V}(T)$ at a small values of $T$, which for $T(t)\approx 1$ implies that, 
\begin{equation}
 3|A|^2+6|A|B-B^2>0,
\end{equation}
where we have taken a negative $A$. It follows that $|A|$ must be bigger that  $-B(1-2\sqrt{3}/2)$ for $\epsilon$ to be much smaller than 1.  In order to assure this small value for $\epsilon$, let us take $|A|^2\geq B^2$. Under this condition, $\epsilon\ll 1$ and therefore, inflation is present if we choose $A$ and $B$ such that $|A|/B\geq 1$, with a negative $A$. The same  conditions assure that $\eta\ll 1$ as well.



\end{document}